\documentclass[11pt]{article}

\usepackage{deauthor,times}
\usepackage{color}
\usepackage{xcolor}
\usepackage{tikz}
\usetikzlibrary{arrows.meta, positioning, shapes}
\usepackage{tikzpeople}

\usepackage{graphicx} 

\title{LLM-Powered Proactive Data Systems}
\author{Sepanta Zeighami, Yiming Lin, Shreya Shankar, Aditya Parameswaran\\
UC Berkeley\\
\{zeighami, yiminglin, shreyashankar, adityagp\}@berkeley.edu
}

\begin{document}

\newcommand{\sep}[1]{\textcolor{blue}{SZ:#1}}
\newcommand{\agp}[1]{\textcolor{red}{Aditya:#1}}

\newcommand{\shreya}[1]{\textcolor{purple}{[Shreya: #1]}}
\newcommand{\yiming}[1]{\textcolor{orange}{[Yiming: #1]}}
\newcommand{\topic}[1]{\smallskip
\noindent {\bf #1.}}

\renewcommand{\thefootnote}{\fnsymbol{footnote}}
  \global\setcounter{footnote}{0}
  \footnotetext{%
    \emph{Copyright {\@BulletinYear} IEEE. Personal use of this material is
    permitted. However, permission to reprint/republish this material for
    advertising or promotional purposes or for creating new collective
    works for resale or redistribution to servers or lists, or to reuse
    any copyrighted component of this work in other works must be obtained
    from the IEEE.}\\
    \textbf{Bulletin of the IEEE Computer Society Technical Committee on
    Data Engineering} \\
    \footnoterule
  }%

\maketitle

\begin{abstract}
With the power of LLMs, we now have the ability to
query data that was previously impossible to query,
including text, images, and video.
However, despite this enormous potential,
most present-day data systems that leverage
LLMs are {\em reactive},
reflecting our community's desire to map LLMs
to known abstractions. 
Most data systems treat LLMs as an opaque black box 
that operates on user inputs
and data as is,
optimizing them much like any other 
approximate, expensive UDFs, in conjunction
with other relational operators.
Such data systems do as they are told, 
but fail to understand and leverage 
what the LLM is being asked to do (i.e.
the underlying operations, which may be error-prone),
the data the LLM is operating on (e.g., long, complex documents),
or what the user really needs. 
They don't take advantage 
of the characteristics of the operations and/or 
the data at hand, or ensure correctness of results when there 
are imprecisions and ambiguities. 
We argue that data systems instead need to be {\em proactive}:
they need to be given more agency---armed with the power of 
LLMs---to understand and rework the user inputs and the data 
and to make decisions on how the operations 
and the data should be represented and processed. 
By allowing the data system to parse, rewrite, and decompose user 
inputs and data, or to interact with the user in ways
that go beyond the standard single-shot query-result paradigm, 
the data system is able to address user needs more efficiently and
effectively. 
These new capabilities lead to a rich design space where the data system takes more initiative: they are empowered
to perform optimization based on the transformation operations, 
data characteristics, and user intent. We discuss various 
successful examples of how this framework has been and can be applied in 
real-world tasks, and present future directions for this ambitious research agenda.    
\end{abstract}

\section{Introduction}

The database community has long acknowledged
the need to store, process, and query data
in various degrees of structure,
from relational, to semi-structured, and more recently,
to unstructured data, including text, images, and video.
Recent developments in AI models,
and LLMs in particular, 
have unlocked the ability to better process
and make sense of unstructured data, in addition to structured data, for tasks
including information extraction~\cite{lin2024towards,liu2024declarative}, summarization~\cite{fang2024multi},  data 
cleaning~\cite{Narayan2022CanFM}, 
dataset search~\cite{huang2023fast}, and data integration~\cite{kayali2024mind}.
LLMs also enable us to better understand 
users, 
manifesting in rapid progress in 
benchmarks on translating natural language 
queries into SQL~\cite{li2024dawn,floratou2024nl2sql}.
LLMs truly have the potential to
disrupt our entire field~\cite{llmsdisruptdatamanagement}.

All of this progress in harnessing LLMs for 
data management---for processing both 
unstructured and structured data---is 
valuable. However, our belief is that we 
are still not leveraging the full potential 
of LLMs for data management.
In most data systems
that leverage LLMs for data processing, 
including those proposed in recent 
work~\cite{patel2024lotus,anderson2024designllmpoweredunstructuredanalytics,liu2024declarative},
LLM operations
are treated as a given, 
i.e., as black-box invocations
on monolithic user inputs and data, where,
akin to other types of UDFs, the data system
doesn't attempt to fully understand the
underlying data, user intent, or constituent operations, 
and just does as they are told.
We call such data systems {\em \bf reactive},
in that {\em they passively execute user-specified 
operations, without understanding 
the underlying intent,
the semantics of the operations, 
or the data on which it should be applied}. 
Reactive data systems 
are fundamentally limited in their ability
to accurately and efficiently address
user needs. 
If the LLM operations
as expressed in the user query have low accuracy
or throw an error, reactive data systems
will faithfully pass the burden of low accuracy or
errors back to the user,
without attempting to proactively 
correct for this.
To understand the limitations of reactive database systems, 
consider the following example.

\begin{example}\label{ex:police}\textit{(Police Misconduct)}
     At UC Berkeley, we are co-leading an effort, along with journalists and public defenders, to build a state-wide police misconduct and use-of-force database\footnote{https://bids.berkeley.edu/california-police-records-access-project}. 
     As part of this effort, our collaborators have gathered, through public records requests, millions of documents detailing
     incidents across 700 agencies.
     Each incident can be split across multiple files, and can
     name several officers.
      Each file itself
     can have many sub-documents, including
     officer testimonies, medical examiner reports, 
     eyewitness reports, and internal affairs determinations.
     The officers
     themselves may be part of several incidents.
     Journalists and public defenders are 
     interested in both investigating the behavior of 
     individual officers,
     as well as broader systemic patterns,
     all in an effort to ensure greater accountability. 
     In such a setting, a reactive data system would 
     encounter various difficulties, as follows:
     \begin{itemize}
         \item \textit{(Difficulties with the Data)}.
         Suppose a journalist is interested in understanding
         the medical impacts of use of force. This is typically 
         detailed in the medical examiner report within
         the broader use-of-force document. 
         Simply providing the LLM the entire 
         document as is (often hundreds of pages) 
         can lead to the LLM making errors~\cite{liu2024lost};
         instead, by decomposing the document into specific semantically meaningful portions
         and focusing LLM attention on those portions
         can both improve accuracy and reduce cost. 
         Another alternative is RAG (Retrieval-Augmented Generation) 
         on pages or chunks, but RAG once again doesn't try to proactively identify 
         the meaning of the documents, or chunks, 
         leading to low accuracies. 
         Here, {\em treating unstructured data as a black box 
         monolith}, as in present-day reactive systems,
         is problematic.

         \item \textit{(Difficulties with the Operations)}.
         The journalists have identified dozens of fields
         of interest in the incidents, including, but not limited
         to: dates, people mentioned, locations, use of firearms, drug use, use of batons and K9 units, among others. 
         Some fields are dependent on other fields,
         e.g., whether there was disciplinary action is contingent on whether there was an internal affairs
         investigation. 
         Simply specifying all of the fields to be extracted
         as is in a single prompt (as a map operation
         or equivalently, a projection)
         can lead to the LLM making errors
         on some of them;
         instead, by decomposing this operation into smaller 
         ``well-scoped'' operations, we can ensure 
         greater accuracy of LLM outputs. 
         Here, {\em treating the operations as a black box},
         without understanding their semantics,
         as in present-day reactive systems, is
         problematic.

         \item \textit{(Difficulties with the User Intent)}. 
         Suppose a journalist is interested in exploring the documents for mentions of a specific officer, ``John Smith''. 
         While a reactive data system would faithfully return
         mentions of John Smith, if any, it would omit
         mentions of officers where the first name is an initial,
         i.e., ``J. Smith'', as well as mentions where
         the middle initial is present, e.g, ``John M. Smith''.
         One could certainly change the  query by requiring a semantic match 
         instead of an exact match---but the journalist 
         would have no way of knowing that such mentions
         exist in the first place.
         A better approach would be to provide, as feedback
         to the journalist, what the query does not currently cover (but could), 
         so that they can make a more informed choice
         about what it is they are actually after.
         Here, {\em treating the user intent as given}, as
         is done in present-day reactive data systems, is problematic.

\end{itemize}
\end{example}
In all three instances, we find that present-day data systems,
especially those that harness LLMs
to help make sense of unstructured data, 
are reactive:
they treat the data, user query, and operations as black-box
indivisible monoliths.
Instead, we argue that data systems should be {\em proactive}:
rather than treating LLM invocations on data as a given, 
such data systems should {\em posess 
the agency to understand user intent, transformation operations, 
and the underlying data}---and to make decisions
on how to best reconfigure the data, operations, and user
input to suit the analysis need.
In the example above, this may include,
for example, 
uncovering underlying layout or patterns 
in unstructured documents,
decomposing (or fusing) operations into semantically
equivalent but more accurate ones,
or going above and beyond immediate user input
to determine the actual underlying user intent,
in concert with the user. 
We argue for {\em truly harnessing the power of LLMs,
to understand and make sense of 
both structured and unstructured data,
rather than simply treating them as black box unstructured data processors}.
We believe the three axes of understanding (1) user intent, (2) data operations 
and (3) the data 
itself are key to the data systems' 
ability to accurately and efficiently processes structured and unstructured data. 

In the following, we will put forth our
vision for {\em proactive data systems---systems
that more effectively harness LLMs for structured and unstructured data
processing by improving our understanding
of data, intent, and operations.}
While our vision is ambitious and expansive,
our early work has already shown promise:
\begin{itemize}
    \item
Our work has shown how
understanding unstructured document collections better, especially
those that obey similar templates, can pay
rich dividends in both cost and accuracy for
document processing~\cite{lin2025twix,lin2024towards}.
\item
Our work 
has shown how a better understanding of error-prone
LLM-based unstructured data processing operators,
as well as the ability to decompose or rewrite
these operators can lead to data processing
pipelines that are a lot more accurate~\cite{shankar2024docetl,parameswaran2023revisiting}.
\item
Our work has also shown
that tailoring our responses
to the underlying user intent,
especially as part of a dialog with the user, 
rather than just strictly adhering to the user
request as stated,
can be very helpful, as evidenced in tasks
that range from data visualization
to dataset search~\cite{zeighami2024nudge,li2024inferring,hulsebos2024took}.
\end{itemize}
Our experience is grounded in our police
misconduct analysis
application, as well as our other work
in understanding where
and how LLMs go wrong, and how we may
be able to avoid these mistakes~\cite{shankar2024spade,shankar2024validates}.

By moving beyond simply executing user
instructions ``as is''
and treating LLM invocations as a black box,
the effective offline and online
execution space of proactive data
systems is effectively unbounded
and open-ended. 
For example, a user task
for extracting information 
from documents can be decomposed
in a potentially unbounded number of 
different ways, with different
accuracies.
Simply leveraging an LLM to, in turn,
do this query planning and optimization for us
can lead to suboptimal results.
Instead, in this vision paper,
we discuss various recipes for
proactive systems to
help make sense of
each of our three axes of data, intent, and operations,
such as performing decompositions 
and rewrites to improve accuracy when performing a given operation, 
finding structure in unstructured data 
to better understand the data 
and answer queries on it, 
and by adjusting data and 
query representations based 
on user feedback to better 
align with user intent.  
We discuss future directions 
along each axis to build better proactive database systems.

\begin{figure}
    \centering
    \includegraphics[width=0.8\textwidth]{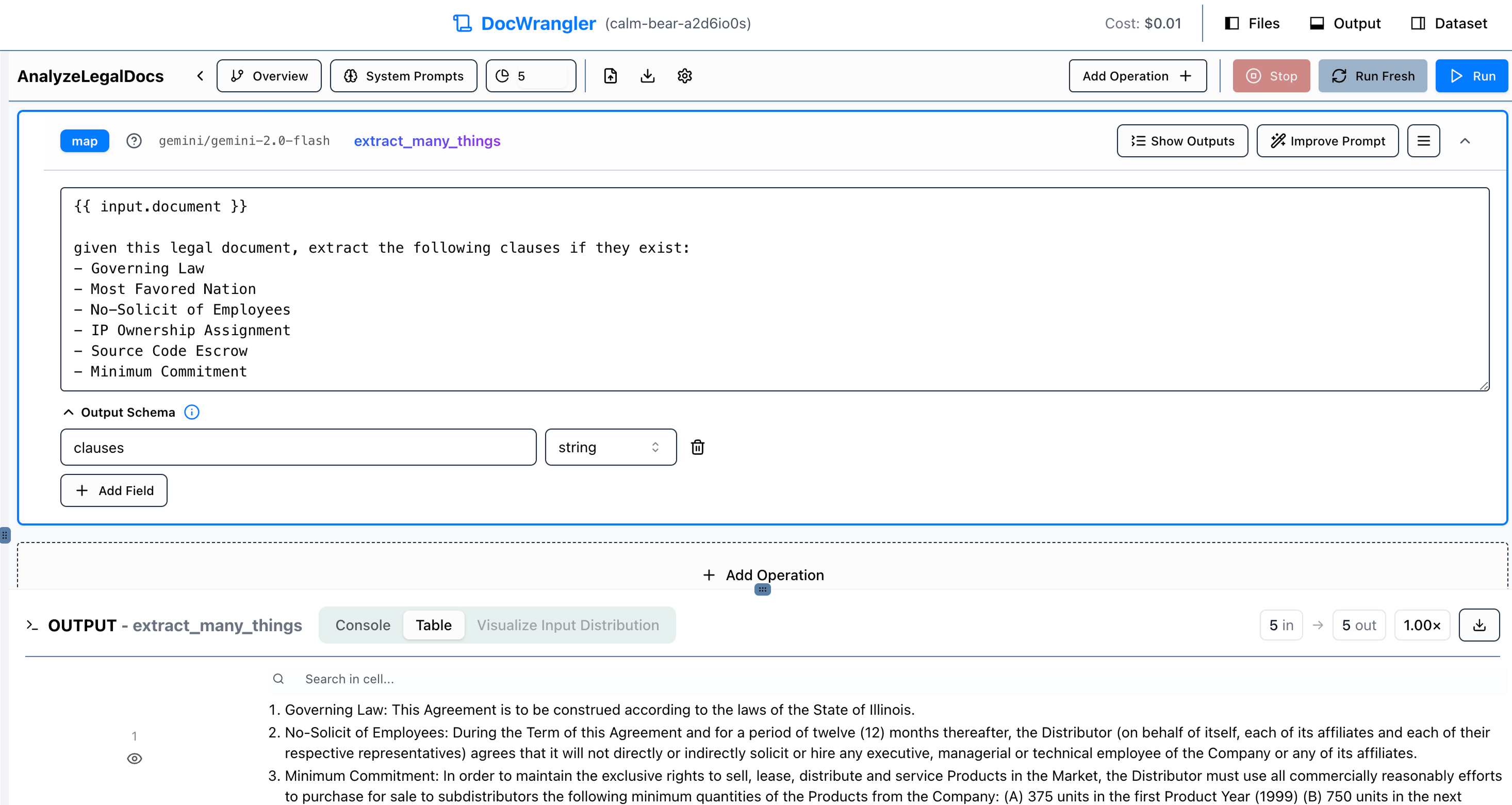}
    \caption{A potential interface for the data system. The user provides a task to be performed on a set of documents, here a collection of legal documents.}
    \label{fig:docetl}
\end{figure}

\begin{figure*}[t!]
\centering
\begin{minipage}{0.55\textwidth}
    \centering
    \includegraphics[width=\textwidth]{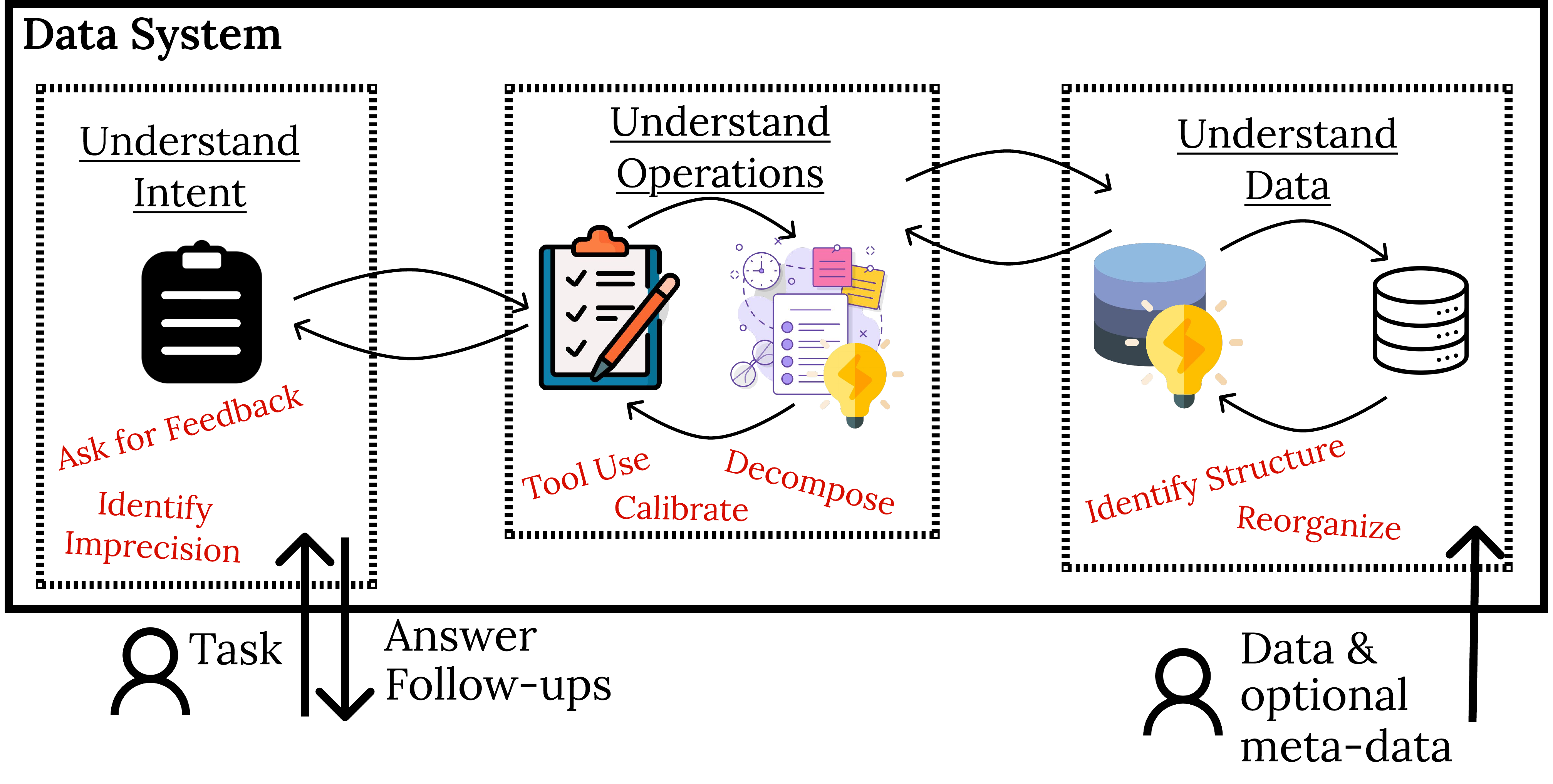}
    \vspace{-15pt}
    \caption{A Proactive Data System (red = LLM-powered)}
    \label{fig:proactive}
\end{minipage}
\begin{minipage}{0.4\textwidth}
    \centering
    \includegraphics[width=0.7\textwidth]{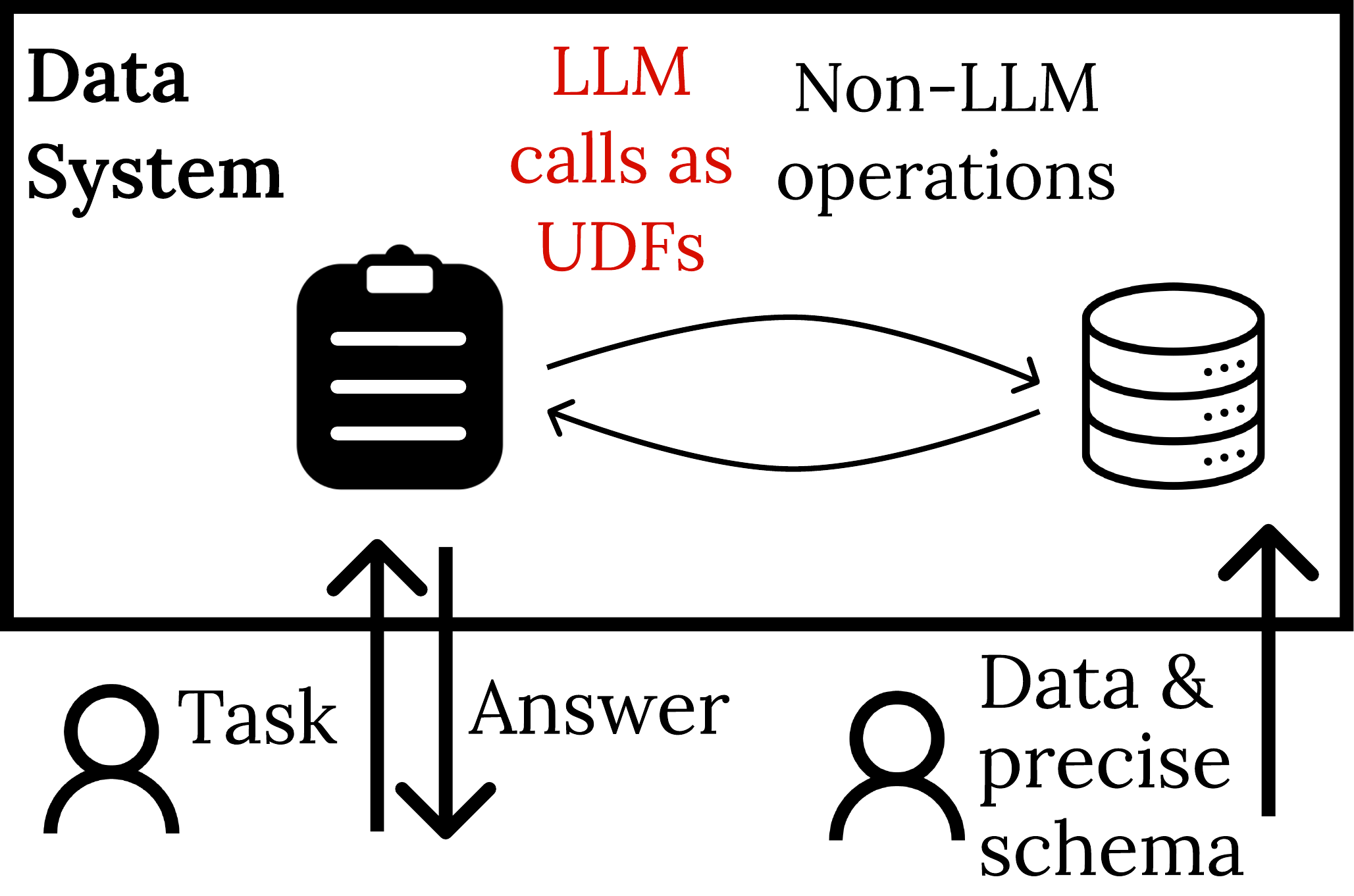}
    \vspace{-10pt}
    \caption{A Reactive Data System}
    \label{fig:reactive}
\end{minipage}
\vspace{-10pt}
\end{figure*}

\section{Typical User Workflows with Proactive Data Systems}

A proactive data system
understands, at a deeper semantic level, 
user intent,
the specific operations it performs,
and the data it performs operations on.
Typical user workflows with 
such a system are similar to
traditional database systems,
where the user
first provides (or ingests)
the data, potentially
alongside additional descriptive
information about its content or schema,
and then proceeds
to execute queries on this data.
We describe this workflow in more detail next,
while acknowledging that
a range of design choices may all
be appropriate,
depending on the use cases. 

\topic{Data Definition} 
The user first ingests
or registers their data (e.g.,
a PDF document collection of incident reports, police officer employment CSVs,
as well as audio/video of incidents 
in the police misconduct case),
along with metadata if any.
Unlike traditional database systems,
this metadata is optional,
and the system
is free to proactively understand the 
structure and semantics of the data.
That said, any 
user-provided specification or description
can help improve
accuracy and better specialize the
system for specific use-cases
of interest.
In Example~\ref{ex:police}, 
such a specification can be similar to 
the first paragraph of Example~\ref{ex:police} as plain text, 
or could include definitions about technical terms or
additional background providing domain knowledge,
e.g., defining police misconduct. 
We note that such information is often needed to allow 
users to query the data meaningfully 
even in relational databases
when using text-to-SQL~\cite{li2024can}. 
We also envision that
in certain use-cases, the users
may proactively identify the entities
of interest for downstream querying,
even if they don't register
the attributes they may care about in the future.
For example, in the police misconduct setting,
the users may want to indicate
that they intend to analyze information
about incidents, police officers,
and agencies.
Armed with all of this information, 
the system
can proactively add indexes,
reorganize the data, 
and extract
certain fields,
among other such offline actions,
as we will describe in Section~\ref{sec:data}.

\topic{Task Specification}
Users are then free
to issue queries or tasks on the data,
which can be specified
in natural language,
or a combination of natural language prompts
associated with data processing operators,
as shown in Figure~\ref{fig:docetl}
for a map operation on a collection of contracts,
with a prompt
extracting a number of legal clauses,
within DocWrangler, our IDE for DocETL~\cite{shankar2024docetl}.
If the user instead chose to preregister
a schema during the data definition
stage, they are free
to extend it with additional
LLM-populated attributes
and issue SQL queries
based on these lazily populated attributes,
as we do in ZenDB~\cite{lin2024towards}.

\topic{Task Execution} 
The system then performs the task on the underlying data.
Rather than performing the task
as is on predefined data monoliths,
a proactive data system
will try to understand
the task and the data,
performing reformulations
of this task by decomposing 
the corresponding operators,
as well as the data,
all in an effort to maximize
accuracy and minimize cost.
For example, the system can decompose 
the user-provided task 
into smaller easier-to-do operations, 
and can leverage the semantics of the document(s) 
to intelligently focus the LLM's attention on 
relevant portions.

\smallskip
\noindent In Section~\ref{sec:op} 
we discuss various ways for a proactive 
data system to understand 
user-defined operations and reformulate them, 
and we extend the discussion in Section~\ref{sec:data} 
on how the system can similarly understand the data 
and perform data transformations to improve accuracy. 
In Section~\ref{sec:intent}, we discuss how 
the data system can ensure the task 
was performed as the user intended.   
When it comes to unstructured data, we
focus our attention mostly on documents
for concreteness,
though our general approach may be applicable
to a variety of formats.

\smallskip
\noindent
To further differentiate a proactive data system 
from a reactive one, Figs.~\ref{fig:proactive} 
and \ref{fig:reactive} provide a high-level overview 
of different components in these systems. 
While a reactive data system considers tasks and data as is, 
and performs operations on the components as instructed, 
a proactive data system leverages LLMs to understand 
user intent, 
the operations it performs, as well as 
the data to ensure maximal accuracy at minimum cost.

\section{Proactive Operation Understanding}\label{sec:op}
In reactive data systems, the onus
is on the user to author queries involving
the ``right'' LLM-powered operators,
with the system then determining how to execute
these in conjunction with other relational operators.
However, even in cases where users
are able to specify clear, unambiguous operations,
the granularity at which they specify them may
not be optimal for execution.
Fundamentally, this stems from a lack
of understanding of what LLMs can do well
versus what they can't---something most users
are not aware of.
In this section, we discuss various 
approaches to 
operation reformulation 
that can improve accuracy 
while maintaining or reducing computational costs.
We first describe new operators that
we may introduce, and then methods for 
assessing and improving cost and accuracy when leveraging
new or existing operators.

\subsection{How and Where to Introduce New Operators}
We now describe ways to reformulate existing LLM-powered
operations into different ones. 

\topic{Decomposition into simpler LLM operations}
In Example~\ref{ex:police}, we described
how sometimes journalists want to extract
dozens of fields from a given document
(e.g., police officer names, descriptions
of misconduct, locations, use of firearms, among others).
The journalist may specify the entire list of fields
along with instructions within a prompt.
However, executing this as is in a single LLM call
may lead to poor accuracies
as LLMs often struggle to identify multiple concepts simultaneously~\cite{shankar2024docetl}.
A proactive system can decompose such an operation
into separate focused ones that
each extract one type of information, improving
accuracy.
An LLM can be asked rewrite a prompt that
says ``extract fields f$_1$, ..., f$_n$ from the following ...''
into ``extract field f$_i$ from the following ...''.
In prior work, we have identified
several new decomposition-based rewrite rules~\cite{shankar2024docetl}
that can lead to higher accuracies,
when coupled with LLMs being used to instantiate
the rewrites themselves.
Recent work on Text-2-SQL also leverages 
similar ideas~\cite{pourreza2024chase}: 
rather than attempting translation 
in ``one-shot'' (i.e., single LLM call), 
which often fails due to schema mismatches 
or poorly-named schemas~\cite{luoma2025snails}, 
it is beneficial to 
parse the operation into smaller units, e.g.,
given ``find all employees who joined before 2020'', 
first identify the core concepts: employee records 
and hire dates,
and treat each separately.

\topic{Leveraging non-LLM components}
In certain cases, operations
that are assigned to an LLM may be better done
through other means, e.g., through SQL.
For example, finding the average settlement amount 
for misconduct cases can be decomposed 
into an LLM operation to 
identify relevant cases and their settlement amounts, 
followed by a SQL aggregation to compute the average.
We have employed similar techniques where
we separate operations that require real-world reasoning (suited for LLMs) from mathematical computations (better handled by traditional database engines or calculators) in
our work on ZenDB~\cite{lin2024towards}.
Determining how to do this automatically
is challenging. 

\topic{Leveraging reasoning or data feedback for reformulation}
As described avove, a proactive data system
must reformulate (e.g., decompose or rework) operations to 
execute them better. 
However, finding good reformulations is difficult. 
Current approaches to discovering good reformulations 
or decompositions
are quite naive. Systems like DocETL~\cite{shankar2024docetl} 
simply prompt LLMs to suggest rewrites, 
either taking the first 
suggestion or selecting from multiple candidates. One option is 
to use a powerful ``reasoning'' model like OpenAI's o1 model to 
rewrite the task, but this still fundamentally relies on one-shot 
prompting, and is unaware of how the rewrite will actually perform on the data. We need approaches that can learn what 
characteristics of the data make tasks challenging, what types of 
LLM errors occur in different contexts, and use this knowledge to 
guide reformulation---perhaps even in an agentic fashion.

\topic{Calibrating LLM outputs}
Independent of which decomposition or reformulation is used,
when LLMs are independently being applied
to a set of items (documents, tuples),
the outputs can often be
inconsistent and non-calibrated.
For example, if we ask LLMs to rate
the severity of every incident
in our document collection,
it often gives all of them the same score,
or worse, gives them scores that
are only losely correlated with the severity.
To remedy this lack of consistency,
we can take various actions.
We can leverage the LLMs themselves to pick 
representative examples that indicate
the full range of the categories of interest, provided
as few-shot examples.
Or they can rework the prompt to describe in more detail
the criteria used for evaluation---to ensure consistency.
Finally, they can also
restrict the space of possible outputs
(e.g., when LLMs are asked to extract state information
from a collection of US addresses, they may extract CA in some
cases and California in others).
While this doesn't change the semantic meaning of the operation, it can significantly improve LLM accuracy by providing better context and guidance, as has
also been explored in work on prompt optimization~\cite{khattab2023dspy}.

\subsection{Assessing and Improving Performance with Reformulation}
Next, we describe ways to 
assess the benefits of reformulations,
and reduce cost while preserving accuracy. 

\topic{Leveraging LLMs to assess benefits}
One question that naturally emerges when considering
decomposing operations into smaller units
is how to assess the benefits of such decompositions.
While it is a-priori hard to tell whether
a decomposition will help, we can run
both the non-decomposed and decomposed
variants on a sample. 
LLMs are much better at evaluating outputs
than generating them, so LLMs can be used
to tell which version performs better.
For example, 
one can employ a ``generate-fix \& rewrite-verify'' pattern: 
generate an initial operation formulation, 
verify its correctness (e.g., through automated checks or LLM 
verification), and if verification fails, attempt alternative 
formulations~\cite{chung2025long}. 
This pattern, 
which we also use in DocETL~\cite{shankar2024docetl}, 
allows a system to systematically explore the space of possible 
formulations until finding one 
that passes verification, effectively optimizing for accuracy 
through trial and refinement.
However, doing evaluation in a cost-effective manner remains
a challenge.

\topic{Deferring to cheaper LLMs for the ``easy bits''}
One concern with decomposition is that
it may increase the cost of the overall pipeline,
since one LLM call per document may now become
multiple. 
One way to defray the cost is to couple decomposition
with cost optimization:
for the simpler newly decomposed operations,
we can alternatively use cheaper and smaller
models to handle them, and only use the more expensive model 
for the most complicated operations.
For example, when we're trying to extract many fields
from a police record document, we can use a cheaper
model for extraction of locations and dates, while using
a more expensive model for harder tasks such as
determining the type of misconduct incident.
While the idea of cheaper proxy models isn't new~\cite{kang2017noscope, patel2024lotus},
here, since the space of decompositions is infinite,
and for each decomposition (or sequences thereof), 
we could use different
models and different confidence thresholds, each with 
different cost-accuracy tradeoffs, the problem
becomes a lot more compliacated.
Additionally, unlike previous settings which focused primarily on 
tasks with well-defined accuracy metrics, we now must provide 
guarantees for open-ended generative operations---where quality 
is harder to quantify.

\topic{Expensive predicate ordering, but with synthesized predicates}
For operations that involve subselecting documents
based on certain criteria (all expressed
together in one prompt), we can leverage existing
related work on expensive predicate ordering~\cite{hellerstein1993predicate, raman1999online};
however, in our context, we can introduce an arbitrary
number of new dependent predicates (that are potentially easier
to check and therefore cheaper).
For example, instead of using an expensive model
to examine each police record document to extract
medical impacts to the victims, if any,
we can consider cheaper filters
that are easier to check, for example, 
if the document contains any medical information at all.
This check could potentially be done by a cheaper
model and rule out a substantial fraction of the documents.
Similarly, when decomposing a complex filter like ``find 
incidents involving both use of force and drug use'' into two 
filters, one for ``use of force'' and one for ``drug use,'' the 
system can evaluate the more selective filter first to minimize expensive LLM calls.



\section{Proactive Data Understanding}\label{sec:data}
Proactive data systems take initiative to truly
understand the data, rather than simply treating
it as inputs to opaque UDF (here LLM) calls.
It can leverage
the provided data descriptions,
as well as actual content,
to create representations
that are useful for downstream data
processing tasks.
The system can understand 
each document on its own (Section~\ref{subsec:structure}), 
understand relationships between documents
or portions thereof (Section~\ref{subsec:cross-doc}), 
or preprocess documents based on anticipated future tasks (Section~\ref{subsec:taskaware}).

\subsection{Identifying Semantic Structure 
within a Document}
\label{subsec:structure}


Although documents may appear unstructured, 
they often are semantically structured.
This structure may
be implicit in the text, e.g.,
content in adjoining portions of the text is often related. 
They can also be explicit, e.g., 
tables or figures embedded within a PDF document. 
We discuss how to identify,
extract, and leverage hidden structure 
from unstructured documents. 

\begin{figure}[tb]
    \centering
    \vspace{-20pt}
    \includegraphics[width=0.7\linewidth]{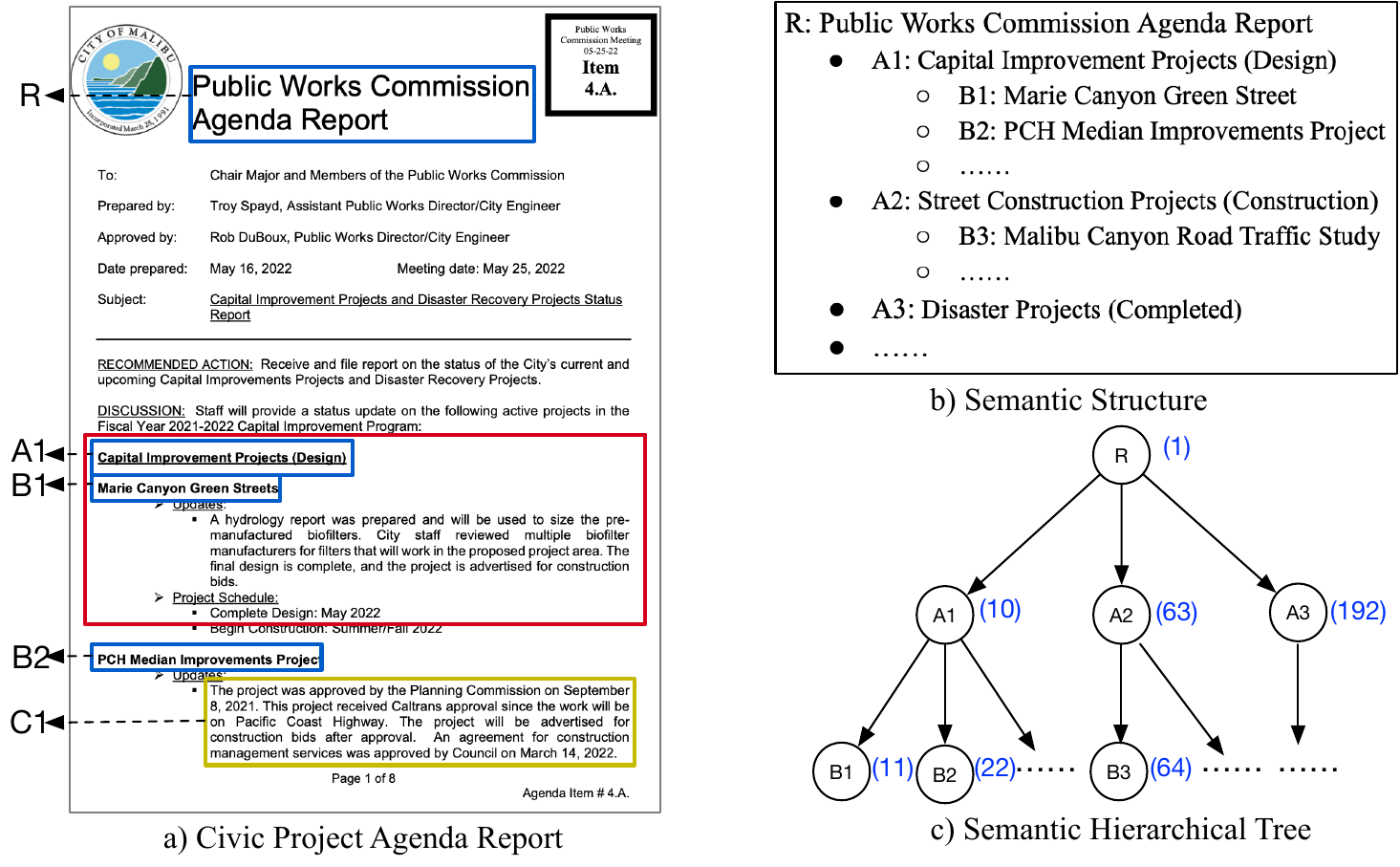}
    \vspace{-10pt}
    \caption{Semantic Hierarchy in a Civic Agenda Document (a) the Document itself (b) the Corresponding
    ``Table of Contents'' (c) the Corresponding Semantic Hierarchy.
    }
    \label{fig:civic}
    \vspace{-10pt}
\end{figure}

\topic{Leveraging implicit hierarchical structure} 
Portions of documents are often semantically related. 
A section or subsection within a document often
contains information that is semantically related,
while other parts are less related or unrelated. 
For example, the medical examiner report within
a broader police record PDF contains
most of the medically relevant information
about an incident, while the eyewitness report
contains most of the relevant information from eyewitnesses.
Identifying these subdivisions within a document
and routing a query to the subdivision at the
``right'' granularity can lead to higher accuracy
than both RAG or providing the entire document to an LLM~\cite{lin2024towards}.
This structure is best represented
as a semantic hierarchy.
There are various ways to construct
such a hierarchy,
including leveraging formatting information
that distinguishes headers from other portions, 
or using an LLM to identify which phrases
may be headers as we do in ZenDB~\cite{lin2024towards}---see 
Figure~\ref{fig:civic} for an example.
Another approach is to construct this semantic hierarchy on
content alone, where summaries of related chunks are merged
and recursively summarized~\cite{sarthi2024raptor}.  
Nonetheless, building semantic structures 
that are useful for downstream tasks remains a challenge, 
as different views of the document may be useful for different tasks,
 where even simple information such as location can have different connotations. 
 For instance, when organizing police activities in a specific case based on location 
 they occurred in, a user might be interested in geographical location 
 of activities (i.e., at a specific address) 
 while another user might be interested in types of locations (e.g., if the police activity was outdoor or inside). 
 The system needs to consider various possible semantics of the data when identifying the semantic structure. 

\begin{figure}[t]
    \centering
    \vspace{-20pt}
    \includegraphics[width=0.7\linewidth]{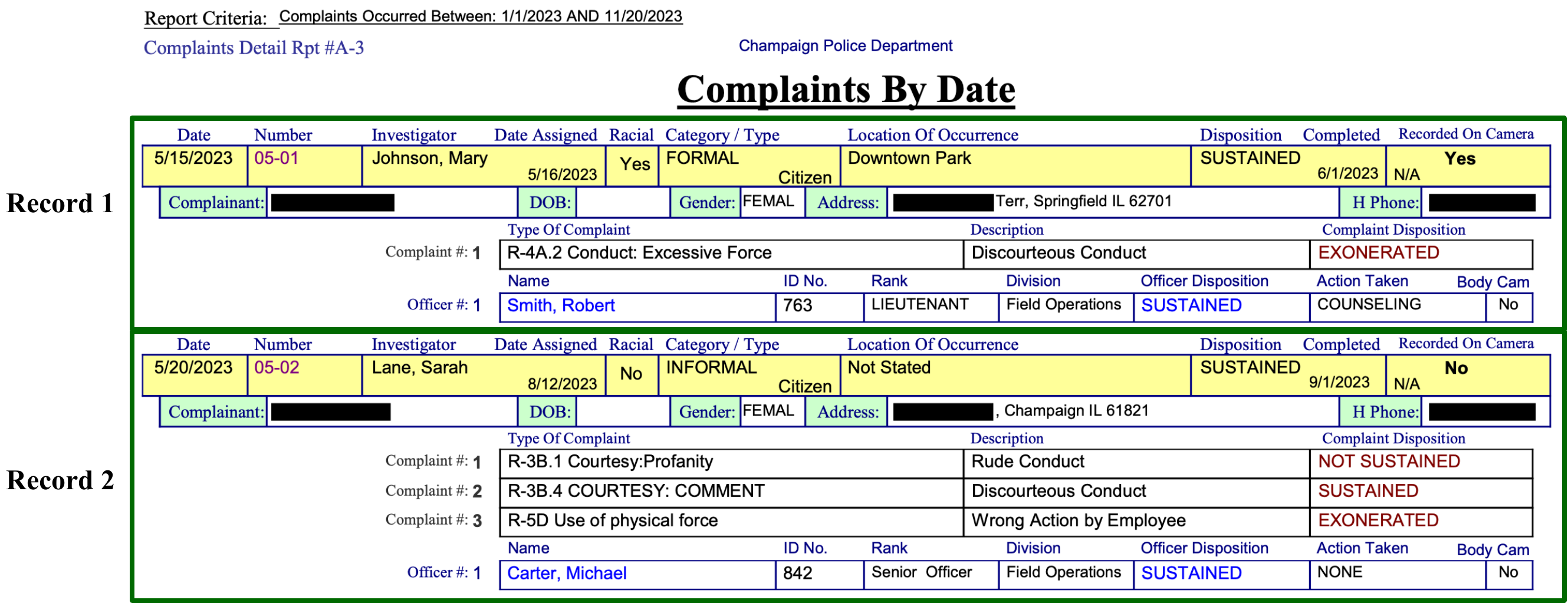}
    \vspace{-10pt}
    \caption{Tables and Key-Value Pairs in Use of Force Records. 
    }
    \vspace{-10pt}

    \label{fig:force}
\end{figure}

\topic{Leveraging explicit structure} 
Unstructured documents often contain structured
portions, such as embedded
tables and key-value pairs.
Treating them as plain text for data processing is ineffective
and error-prone.
For example, if we're not careful in preserving visual information, 
a missing value in a key-value pair
could lead to the next key being misinterpreted
as the corresponding value.
Moreover, depending on the approach used
to query such tables,
we may lose visual information (used to show table structure and group columns and rows),
and be unable to effectively process
numerical information. 
A proactive data system
therefore will identify and extract such structured portions
and represent them in a structured format, for example, as tabular data or key-value pairs in Figure~\ref{fig:force},
while preserving their context within the document (e.g., their location and semantic relationships to the rest of the document).
Our recent tool, TWIX~\cite{lin2025twix}, proposes an efficient approach 
for automatically extracting structured portions from documents, using a combination of visual
and LLM-based inference, while preserving this context for the extracted
information. 
However, many challenges remain, such as accurately representing the semantic 
relationship between structured and unstructured document portions, 
e.g., to understand which queries should be answered 
based on the structured and unstructured portions,
and how much background context is necessary to make sense of the structured portions.

\subsection{Identifying Cross-Document Relationships}\label{subsec:cross-doc}
There are multiple reasons
to perform cross-document organization.

\topic{Identifying documents that may be queried together}
Beyond understanding structure within a single document, 
it is important to understand relationships across documents,
since these related documents may often be queried together. 
In Example~\ref{ex:police}, the dataset, a single incident
can span several PDF documents, often without such information being linked to each other. 
The data system needs to proactively identify relationships 
between such documents to organize the data prior to querying. 
This, for instance, can be done by clustering the documents. 
However, clustering is challenging, since the system needs to understand
the documents to be able to cluster them properly.
Simply embedding the documents, and clustering the embeddings 
does not work since the documents can vary considerably in length.
Another approach is to leverage LLMs to check if two documents
correspond to the same incident, but this is expensive,
especially when there are $O(n^2)$ comparisons.
We may be able to leverage LLMs to identify cheaper proxies
or blocking rules (e.g., two documents
may not be related unless the date ranges overlap)
for this organization. 
In some settings, folder organization provides cues for identifying
cross-document relationships (e.g., documents that are very ``far apart''
from a folder structure standpoint may be unlikely to be related).

\topic{Identifying shared templates across documents}
A separate concern is to combine semantic hierarchy construction
with cross-document relationships, so that we are able
to identify shared ``templates'' across documents.
These templates can both help scale up extraction across
documents, but also help identify documents whose structure
differs considerably.
For example, journalists may want to identify incidents
where there is an internal affairs report within
a broader police record document,
since these are ones where there is a corresponding
disciplinary action. 

\subsection{Task-Aware Data Pre-Processing}\label{subsec:taskaware}
The system can attempt to proactively find and organize portions
of the data that will be useful to improve performance on a reasonable subset of data processing tasks downstream. 
Given that document collections 
can span in the millions, 
it can be expensive to do extensive processing of the data upfront, for instance,
by populating a materialized view with all the attributes a user can ever hope to query;
it can also be  time-consuming to leave all the data processing to when the user issues a task. 
As such the system needs to decide how much preprocessing is beneficial upfront, and what to perform at query time. 
To strike a balance between the two extremes, 
one option is for the system to identify and/or extract data units 
that it deems to be useful in the future for a wide variety of queries. 
This can be done by understanding the semantics of the data. 
For instance, in Example~\ref{ex:police}, the system can decide that sections 
that describe police incidents at a high level (e.g., the internal affairs report)
are typically useful for future task processing as they provide a comprehensive summary of most
relevant aspects. 
The system can keep pointers to such sections as lightweight indexes, 
but leave more specific data processing to when the user issues queries. 
Similarly, the system can do schema identification in advance 
to find what type of data is represented in the documents, 
and use the identified schema to answer queries. 
The system can decide whether to extract information 
upfront to populate the schema, 
or keep pointers to where the information 
can be found at query time. 
For instance, the system may choose to retain pointers
to all portions that mention police officers in the document
so as to accelerate analysis of those aspects downstream,
without going all the way to populating a materialized
view with officer attributes (since these can vary 
depending on user need).

\section{Proactive User Understanding }\label{sec:intent}

Even with best-effort operation reformulation 
and data understanding capabilities, 
a fundamental challenge remains: the gap between what users specify 
and what they actually need. 
This challenge manifests in multiple ways---users may provide ambiguous specifications, 
fail to articulate implicit assumptions, 
or simply not know how to express their requirements fully~\cite{papicchio2024evaluating, shankar2024validates}. 
To bridge this gap between the user's intent and the operations performed, 
a proactive data system needs to be internally 
aware of this gap when executing the task (Section~\ref{sec:internal}), 
leverage user feedback to bridge the gap (Section~\ref{sec:feedback}) 
and provide mechanisms for users to externally validate query results (Section~\ref{sec:verification}). 

\subsection{Imprecision-Aware Processing}\label{sec:internal}

Unlike reactive data systems
that only execute the query as stated, proactive
data systems can leverage LLMs to help truly
identify user intent,
despite the user-provided tasks representing
an imprecise or incomplete specification thereof. 
The system should therefore internally consider 
multiple possible user intents when processing tasks, 
potentially providing different answers that correspond 
to these different interpretations. This can be done to varying degrees. 
In the simplest form, the system can consider multiple interpretations 
of the statements provided by the user. 
In Example~\ref{ex:police}, the system can consider 
various spellings of the same name, 
either in the input provided by the user or derived from the data. 
Such attempts are similar to 
possible world notions in the database community~\cite{suciu2022probabilistic}, 
where the database can consider various possibilities for ``fuzzy'' data and queries.

A proactive system can take 
more aggressive steps to understand user intent. 
For example, the system can consider adding new predicates 
that the user may be interested in---e.g., 
if a user has previously asked several questions about police misconduct 
in a specific city but submits a new query without specifying location, the system might prioritize results from that city. 
This intent discovery can be data-driven---the system might determine that records from certain cities are more relevant 
or interesting 
and prioritize them in the output. 
Moreover, if the output for an operation is too large, the system can 
selectively display what it determines to be the most relevant answers 
or provide appropriate summaries or sample outputs.

The system can also anticipate user questions, for example, ``why was a certain record not provided in the answer'', 
and proactively relevant records that, 
while not strictly matching the query, might be of interest to the user. 
Additionally, the system can modify queries by dropping or relaxing certain predicates---e.g., if the user has specified a predicate that leads to empty results. 
Or, the system might expand query scope, for example, geographically, to include potentially interesting results 
(such as when a specific type of police misconduct, while not present in the queried city, occurred in neighboring jurisdictions).

\subsection{Leveraging User Feedback}\label{sec:feedback}
A proactive system can leverage feedback from users to clarify intent in a lightweight manner.
This feedback serves two purposes: improving accuracy for the current task, and enhancing the system's understanding of user intent for future queries.

To improve the accuracy on the current task, the system can decide to ask follow-up questions~\cite{li2024llms}. This might include asking for clarification about task goals, gathering additional specifications, or presenting example results for users to indicate which best match their needs. The important challenge is balancing the need for clarity with minimizing user burden: for example, when processing police misconduct documents, rather than asking multiple detailed questions, the system might show representative document types and let users select which are most relevant---then apply this learning broadly across the document collection. Similarly, when encountering potential name variations in Example~\ref{ex:police}, the system might ask a single question about handling typos that can inform its overall matching strategy, rather than asking the user to confirm every typo correction. While LLMs offer promising capabilities for generating targeted feedback requests, automatically determining what feedback to request and when remains an open research challenge. Prior work
on predictive interaction is highly relevant here~\cite{2015-predictive-interaction}.

User feedback can also be leveraged to improve system performance on {\em future} tasks. For example, if the user provides feedback that a certain document is relevant or not relevant to a task, the system can then update its data and task processing mechanism to take that feedback into account. This can, for instance, change operation rewrite rules used internally for processing (Section~\ref{sec:op}), modify data representation~\cite{zeighami2024nudge} or change how semantic structure is extracted from the data (Section~\ref{sec:data}). While this approach shares similarities with query-by-example systems that learn from user-provided examples, e.g., \cite{fariha2018squid}, it extends the concept more broadly—--allowing the system to refine its understanding of user intent across a diverse range of tasks and feedback types.

\subsection{Verifying Execution}\label{sec:verification}

A proactive system must provide users with the means to verify that their tasks were executed correctly. Verification is particularly important when the system makes autonomous decisions---for instance, when correcting potential typos in names, the system should clearly show which corrections were made to allow users to catch any incorrect modifications. 
If, during processing, the system encountered anomalous documents, it's best to indicate them as such to the users so that they don't pollute the rest of the analysis.

While the system can provide comprehensive execution traces, including details of LLM operations performed and data sources accessed~\cite{tan2007provenance}, presenting this information in a user-friendly way remains challenging. Simply showing raw execution traces or complete datasets is overwhelming and impractical, as users cannot reasonably review large amounts of data to verify correctness. An interesting open challenge
is to determine a small subset or explanation that conveys the same information as the entire provenance; we can always verify such explanations using an LLM.

\section{Conclusion}

The database community stands at a pivotal moment where LLMs offer unprecedented capabilities for processing both structured and unstructured data. In this vision paper, we proposed {\em proactive} data systems: systems that possess agency in understanding and optimizing data processing tasks. Unlike traditional {\em reactive} systems that treat LLMs as black-box UDFs operating on monolithic inputs, proactive systems go further in leveraging LLMs to aid data processing along three axes. We presented these axes---operations, data, and user intent---and demonstrated the potential of LLMs to help in each one through our recent work on operator reformulation, document organization and analytics, and intent-aware optimization. Overall, proactive data systems can achieve both higher accuracy and lower costs than reactive systems that treat LLMs as black boxes.


{\scriptsize
\bibliographystyle{ieeetr}
\bibliography{references}
}

\end{document}